\documentclass[12pt]{JHEP3}
\pdfoutput=1

\usepackage{amssymb, amsmath, amsopn, amsthm}
\usepackage{epsfig,amsfonts,bm}
\usepackage{graphics}

\def\bec{\begin{center}}
\def\eec{\end{center}}

\def\p{\partial}

\newcommand{\be}{\begin{equation}}
\newcommand{\ee}{\end{equation}}
\newcommand{\bea}{\begin{eqnarray}}
\newcommand{\eea}{\end{eqnarray}}
\newcommand{\beas}{\begin{eqnarray*}}
\newcommand{\eeas}{\end{eqnarray*}}

\def\scriptlap{{\kern1pt\vbox{\hrule height 0.8pt\hbox{\vrule width 0.8pt
  \hskip2pt\vbox{\vskip 4pt}\hskip 2pt\vrule width 0.4pt}\hrule height 0.4pt}
  \kern1pt}}

\def\Biggg#1{{\hbox{$\left#1\vbox to 25pt{}\right.\n@space$}}}
\def\n@space{\nulldelimiterspace=0pt \m@th}
\def\m@th{\mathsurround = 0pt}

\title{Towards the Lattice Effects on the Holographic Superconductor}

\author{Norihiro Iizuka$^{1}$ and Kengo Maeda${}^2$
\\

{\small \sl ${}^1$Theory Division, CERN, CH-1211 Geneva 23, Switzerland} \\
{\small \tt  {norihiro.iizuka@cern.ch}} \\

{\small \sl ${}^2$Faculty of Engineering,
Shibaura Institute of Technology,} \\ 
{\small \sl Saitama, 330-8570, Japan} \\
{\small \tt {maeda302@sic.shibaura-it.ac.jp}}

\vspace{0.1cm}

}

\abstract{
We study the lattice effects on the simple holographic toy model; 
massive $U(1)$ gauge theory for the bulk action. 
The mass term is for the $U(1)$ gauge symmetry breaking in the bulk. 
Without the lattice, the AC conductivity of this model shows similar 
results to the holographic superconductor with the energy gap. 
On this model, we introduce the lattice effects, 
which induce the periodic potential and break the translational invariance of the boundary field theory.  
Without the lattice, due to the translational invariance   
and the mass term, 
there is a delta function peak at zero frequency    
on the AC conductivity. 
We study how this delta function peak is influenced by the lattice effects, 
which we introduce perturbatively.  
In the probe limit, 
we evaluate the perturbative corrections to the conductivities at very small frequency limit. 
We find that the delta function peak remains, 
even after the lattice effects are introduced, although 
its weight reduces perturbatively. We also study the lattice wavenumber dependence of this 
weight. 
Our result suggests that in the $U(1)$ symmetry breaking phase, 
the delta function peak is stable against the lattice effects at least perturbatively.

}

\preprint{CERN-PH-TH-2012-197}
\begin{document}

\section{Introduction}\label{sec:intro}

Finding the high $T_c$ superconductors on the cuprates system, 
which are not described by the usual BCS theory,  
is a remarkable breakthrough in the condensed matter physics occurred almost  26 years ago.  
Even though there are huge developments after that, both theoretically and 
experimentally, we are still missing the core mechanism governing the system. 
One of the common mysterious phenomena in these superconductivity is its 
Non-Fermi-liquid behavior above the superconducting phase, which occurs  
near the quantum critical point. 
The main difficulty to understand the mechanism of these high $T_c$ superconductor is, of course, 
its strongly coupled dynamics and its lack of the normal quasi-particle pictures.  
We do not yet fully understand by what mechanism and for what materials, 
how high $T_c$ superconductor can occur in nature. 
However, as is recent discovery of iron-based superconductor, 
experimental progresses on this field are remarkable. 
These include recent developments of 
the cold atom experiments. 
Therefore it can happen that in the near future, 
we get more crucial experimental data which 
helps to deepen our understanding of the core mechanism.

On the other hand, one of the most surprising development coming from the 
string theory is the realization of holographic principle,   
which states that two totally different theories, 
string (or gravitational) theories in asymptotically anti-de Sitter space background 
and strongly coupled large $N$ gauge theories, 
are equivalent at some limit \cite{Maldacena:1997re,Witten:1998qj,Gubser:1998bc}. 
Recent developments of the holography by applying that 
to the strongly coupled condensed matter system, 
is just tremendous\footnote{See for examples, 
\cite{Hartnoll:2009sz, McGreevy:2009xe, Sachdev:2010ch, Hartnoll:2011fn}}. Especially the construction of the holographic superconductor (superfluid) 
\cite{Gubser:2008px, Hartnoll:2008vx,Hartnoll:2008kx}, 
where the $U(1)$ symmetry breaking through the hairy black hole in the bulk, 
intrigues the many interesting developments.  
See, for examples, \cite{Herzog:2009xv, Horowitz:2010gk} for the review of the 
holographic superconductors.

In the real-world materials, it happens quite frequently that the materials showing the 
superconducting phase do not have a translational invariance and Lorentz symmetry, due to the 
crystal structure of the background atoms.  
In high $T_c$ superconductor, the effects of the background atomic structure are  
very important and it is expected that two-dimensional structure plays the significant role. 
One of the very important effects of the background atomic lattice is 
that it violates the translational invariance and Lorentz invariance and induces the 
periodic potential.   

If the material possesses a translational invariance and a net charge, 
it is more or less guaranteed that 
its electric conductivity shows the delta function peak at zero frequency. 
This is simply because of the fact that the charged objects 
are kept accelerated by the outer electric field in a translationally invariant system. 
We can also understand this from the fact that by the Lorentz boost, 
the system acquires a nonzero current with zero applied electric field.  
However, once we break the translational invariance by the lattice, 
then, there is no guarantee that such a delta function peak  appears 
on the conductivity\footnote{Even if the system has a translational invariance, 
if we apply magnetic field, this delta function peak disappears. 
This is because momentum is not conserved in the 
presence of magnetic field. 
The same is true in the presence of disorder. 
See, for example,  \cite{Goldstein:2010aw, Maeda:2010tc} for the explicit examples of these in the holographic conductivity calculations.}. 
Therefore it is quite interesting to consider how the delta function peak  in 
many interesting holographic system is influenced 
by these lattice effects.  
In this paper, we take a first step towards the lattice effects 
on the holographic superconductor; we study the lattice effects on a toy model, 
which is massive $U(1)$ gauge theory for the bulk action. 
The mass term of gauge boson is for the $U(1)$ gauge symmetry breaking in the bulk. 

Our toy model, although it is a different theory from the holographic superconductor model  
analyzed by \cite{Hartnoll:2008vx,Hartnoll:2008kx}, has properties which is similar 
to the holographic superconductor; Without the lattice, it shows the mass gap, and AC conductivity 
is quite similar to the results of \cite{Hartnoll:2008vx}. 
It also shows the delta function peak at zero frequency.  
Therefore, we find it interesting to ask 
how the zero frequency delta function peak in our model, 
which is due to the translational invariance and the mass term,  
is influenced by the lattice effects. 
Since in our model the gauge boson has mass term, corresponding to 
the $U(1)$ symmetry breaking of the superconductor (superfluid) phase, 
this analysis is a first step to study the generic lattice effects on the holographic superconductor 
phase. 

There are several technical points which are worth quoted at this stage. 
In this paper, we consider the probe limit, namely we neglect the effects of the gravity. 
It is known that in the normal phase ({\it i.e.,} non-superconducting phase) without taking into 
account the gravity, conductivity becomes trivial and there is no delta function peak appearing. 
The delta function peak appears only if we take into account the gravity effects in the normal phase. 
On the other hand, in the superconducting phase of  \cite{Hartnoll:2008vx},   
the delta function peak appears without taking into account the gravity effects, so is 
our bulk massive $U(1)$ gauge model. 
In this paper, without taking into account the gravity, 
we study how the delta function peak in our model is influenced if we introduce the lattice effects 
perturbatively (periodic disturbance to the system).

Before we end this introduction, we comment on several closely related references. 
Recently there are developments to take into account the lattice effects for the 
holographic condensed matter system.
In the paper by Maeda, Okamura, and Koga \cite{Maeda:2011pk}, the geometry, 
where the back reaction of lattice effects is taken into account perturbatively, is constructed. 
In that paper, the lattice effects are introduced through the chemical potential. 
In \cite{Horowitz:2012ky},  Horowitz, Santos, and Tong 
calculated the conductivity under the presence of the 
lattice for the {\it normal} phase, namely, non-superconducting phase. 
They showed, using a 
very powerful numerical technique, how the AC conductivity, especially the zero frequency 
delta function peak,  
is influenced by the lattice effects. They showed that the delta function peak disappears 
by the lattice effects, 
 as is expected for the 
properties of the normal phase. 
In that paper, the lattice effects are introduced through the neutral scalar field. 
Very recently, a paper \cite{Liu:2012tr} by Liu, Schalm, Sun, Zaanen, 
appears where on the geometry without gravitational back reaction, they 
discussed the lattice effects for the holographic 
fermion correlators, especially 
 its pole for the Non-fermi-liquids, to see how their dissipation relations are modified.

\section{Massive $U(1)$ gauge boson model}
\subsection{The model}

In this paper we consider the following toy model of holographic superconductor for the bulk action. 
\begin{align}
\label{action:U(1)_gauge}
S=  \int  d^4x \sqrt{-g}
\left(-\frac{1}{4}F^2-V(u,x)A^2\right)  \,,
\end{align}
where $V(u,x)$ is external potential, and plays the role of position-dependent mass term for 
the gauge boson $A_\mu$. $u$ is radial coordinate in the bulk and $x$ is the spatial coordinate in the boundary theory. 
In more realistic holographic superconductor model, $V(u,x)$  is given by the 
condensation of the charged scalar field $\Psi$ as \cite{Gubser:2008px, Hartnoll:2008vx,Hartnoll:2008kx}. 
There, the massless $U(1)$ gauge boson couples to the charged scalar 
$\Psi$, where $\Psi$ takes non-zero VEV, $\Psi^{background}$ 
\bea
\label{nonzeroVEVforchargedscalar}
\Psi = \Psi^{background}(u,x) \neq 0 \,.
\eea
This gives the potential 
\bea
V(u,x) \sim |\Psi^{background}|^2 \,.
\eea
corresponding to the spontaneous $U(1)$ symmetry breaking in the bulk, 
which is dual to the global $U(1)$ symmetry breaking in the boundary theory. 
However in this paper, we consider $V(u,x)$ as given input for the symmetry breaking. 
Especially, we consider $V(u,x)$ which satisfies 
\bea
V(u,x) \to 0  \quad (\mbox{at the boundary}), 
\eea
so that at the boundary, the mass term $V(u,x)$ for the gauge boson disappears. 
We would like to calculate the AC conductivity through the 
bulk $U(1)$ dynamics $A_\mu$, and discuss the delta function peak with the lattice effects.

One of the main reasons why we consider this toy model is its simplicity. 
By restricting the degrees of freedom, the calculation for the AC conductivity, especially 
by choosing appropriate boundary condition, becomes much simpler than the holographic 
superconductor model in \cite{Hartnoll:2008vx,Hartnoll:2008kx}. 
However, as we will show later, this model, without the lattice, 
shows the AC conductivity which is quite similar to the one of the 
holographic superconductor model. It shows the energy gap. 
Since one of the essential features of the holographic superconductor model 
is the $U(1)$ symmetry breaking in the bulk, we expect this bottom-up model 
captures some of the essential features. In this paper, we consider the lattice 
effects on this model.

The equations of motion for gauge field become
\begin{align}
\label{eq:U(1)_gauge}
\nabla_\nu F^{\mu\nu}=\frac{1}{\sqrt{-g}}\p_\nu 
(\sqrt{-g}F^{\mu\nu})=-2V(u,x)A^\mu \,, \quad
F_{\mu\nu}=\nabla_\mu A_\nu-\nabla_\nu A_\mu \,. 
\end{align}
We give, by hand, the non-zero arbitrary VEV for the background gauge potential $A_\mu$ as   
\bea 
\label{nonzeroVEVforgaugeboson}
A_\mu = A_{\mu}^{ background}(u,x) \neq 0 \,, 
\eea
which should be correlated with the nonzero VEV of the charged 
scalar field (\ref{nonzeroVEVforchargedscalar}) in holographic 
superconductor \cite{Hartnoll:2008vx,Hartnoll:2008kx}. This nonzero VEV  
produces a net charge for the boundary theory through the normalizable mode of $A_\mu^{ background}$.  
On this background, in order to calculate the conductivity, 
we will add small fluctuations given by the ansatz 
\begin{align}
\delta A_\mu dx^\mu = \delta A_t (t,u,x) dt+ \delta A_x (t,u,x) dx + \delta A_u (t,u,x) du \,.
\end{align}
From the dynamics of these fluctuations $\delta A_\mu$, we would like to 
calculate the AC conductivity. However, since the $U(1)$ gauge boson equations of motion are  
linear equations, all of our analysis are independent on the VEV of the gauge boson (\ref{nonzeroVEVforgaugeboson}). 
Therefore our argument for the calculations of the conductivity is independent on the 
background VEV (\ref{nonzeroVEVforgaugeboson}).

We take the background metric to be 
\begin{align}
\label{Sch_AdS}
ds_{background}^2=\frac{L^2}{u^2} \left( -h(u)dt^2+dx^2+dy^2 +\frac{du^2}{h(u)} \right) \,, 
\end{align}
where $h$ is radial dependent function, given by  
\bea
h(u)=1-u^3 \,,
\eea
corresponding to the Schwarzschild AdS black brane. 

We study the lattice effects on our  model given by the action (\ref{action:U(1)_gauge}). 
As we quoted already, this model is a different theory from the holographic 
superconductor one \cite{Hartnoll:2008vx}, 
where the degrees of freedom is $U(1)$ gauge boson and the charged scalar field. 
We give the $U(1)$ symmetry breaking background by hand as 
(\ref{nonzeroVEVforchargedscalar}) and (\ref{nonzeroVEVforgaugeboson}). 
We have not taken into account the dynamics of  the charged scalar field fluctuation in the 
$U(1)$ symmetry breaking phase, which might not be consistent from the original 
gauge boson and charged scalar field theory in the bulk. 
Therefore the reader may regard our model a rather toy model akin to the 
$U(1)$ symmetry breaking phase of the holographic superconductor.  
Although our massive $U(1)$ gauge boson model is different model, there are 
several merit to study this model. First, due to the lack of the dynamical charged scalar, the 
analysis of solving the equations of motion is simpler in this model. 
Second, even though we will not take into account the charged scalar field fluctuation 
for the conductivity calculation, the results of conductivity without the lattice in our model 
are quite similar to the 
superconductor model, it shows the mass gap. 
Technically, this is because charged scalar fluctuation does not 
directly couple to the gauge boson fluctuation without the lattice. 
Third, this simple model also shows the zero frequency delta function peak 
in the probe limit of the holographic 
superconductor. Therefore even though this model is different, since this model 
shows very similar properties to the $U(1)$ symmetry breaking phase of the 
holographic superconductor, 
we regard this model belongs to the same category to the one in \cite{Hartnoll:2008vx}.
Therefore even in this model, there are interesting questions we can ask.   
In this paper, we restrict our attention to this model, 
 and we study the lattice effects on it.

We also point out that in this work, we neglect 
gravitational back reaction. We expect that even if we go beyond the probe limit, 
the back reaction does not change the picture drastically as the case of holographic 
superconductivity. 
For this reason, we take the background to be Schwarzschild AdS black brane as (\ref{Sch_AdS}), 
even though there are non-trivial background flux (\ref{nonzeroVEVforgaugeboson}). 
Once we take into the 
back reaction, the metric will be modified either into Reissner Nordstr\"om  AdS type of brane 
or AdS hairy black brane solution. 
 
\subsection{Generic analysis} 
We would like to consider the gauge boson fluctuation so that we can 
obtain AC conductivity. For each of the fluctuation (\ref{eq:U(1)_gauge}), 
the equations of motion becomes 
\bea
\label{eq:At}
h \, \delta A_{t,uu} 
- 
(\delta A_{x,tx} - \delta A_{t,xx})
- h \; \delta A_{u,ut}
-\frac{2L^2}{u^2}V(u,x) \delta A_t &=& 0 \,,  \\
\label{eq:Ax}
h \, \frac{\p}{\p u}(h \, \delta A_{x,u})-   
( \delta A_{x,tt} -  \delta A_{t,xt})  
 - h \frac{\partial}{\partial u} (h \, \delta A_{u,x}) 
-\frac{2L^2h}{u^2}V(u,x) \delta A_x &=& 0 \,, \quad \quad \\ 
\delta A_{t,tu} - h \, \delta A_{x,xu} 
- \delta A_{u,tt} + h \, \delta A_{u,xx} - \frac{2  L^2 h  }{u^2} V(u,x) \delta A_u &=& 0 \,.
\eea
Equation of motion for the $y$ component is trivially satisfied by $\delta A_y = 0$. 

Clearly, if the potential $V(u,x)$ is independent on position $x$, then there exists the solution 
where both $\delta A_t = \delta A_u = 0$ with nontrivial  $\delta A_x$, which is independent on $x$. 
In such cases, 
fluctuation equation for $\delta A_x$ becomes single differential equation for the second order. 

Taking the simple time-dependence 
as $ \delta A_i=e^{-i\omega t}a_i~(i=t,x,u)$ for the AC conductivity, we obtain 
\bea
\label{eq:at}
 ha_{t,uu}+  
  a_{t,xx}+i\omega a_{x,x} 
+ i \omega h a_{u,u}
-\frac{2L^2}{u^2}V(u,x)a_t &=& 0 \,, \\
\label{eq:ax}
 h\frac{\p}{\p u}(ha_{x,u})+ 
 \omega^2a_x-i\omega a_{t,x}
- h\frac{\p}{\p u}(ha_{u,x})
-\frac{2L^2h}{u^2}V(u,x)a_x &=& 0 \,, \\
\label{eq:au}
 - i \omega a_{t,u} - h a_{x,xu} 
+ \omega^2 a_{u} + h a_{u,xx} - \frac{2 L^2 h  }{u^2} V(u,x) a_u &=& 0 \,.
\eea

We would like to solve these coupled differential equations by perturbation 
and obtain the AC conductivity. 
For that purpose, we add the lattice effects which simply take the 
following  cosine form by the perturbation as 
\begin{align}
\label{gauge_potential}
V= 
\frac{1}{L^2} \left( V_0(u) + {\epsilon} \, \delta V(u) \cos qx \right) \,, 
\end{align}
where $\epsilon$ is a small parameter. 
$V_0(u)$ corresponds to the homogeneous charged scalar condensation, while 
$\delta V(u)$ corresponds to the lattice effects. We take the ansatz that 
the lattice has position $x$ dependence given by the wavenumber $q$.   
We will take an explicit example later, but for a moment we keep it generic $u$-dependent 
functions $V_0(u)$ and $\delta V(u)$.

We will conduct perturbation expansion for small $\epsilon$ as 
\begin{align}
\label{expansion}
& a_x=a_x^{(0)}(\omega, u)+\epsilon a_x^{(1)}(\omega, u) \cos qx+
\epsilon^2a_x^{(2)}(\omega, u, x)+\cdots, \\
& a_t= \, \epsilon a_t^{(1)}(\omega, u) \sin qx+
\epsilon^2 a_t^{(2)}(\omega, u, x)+\cdots, \\
& a_u=\epsilon a_u^{(1)}(\omega, u) \sin qx+
\epsilon^2 a_u^{(2)}(\omega, u, x)+\cdots \,.
\end{align}
Note that $a_i^{(0)}$ and $a_i^{(1)}$ are independent on $x$ but we keep 
the implicit $x$ dependence for $a_i^{(2)}$. 
Then, from (\ref{eq:at}), (\ref{eq:ax}), (\ref{eq:au}), we obtain  
\begin{align}
\label{eq:a_x0}
& h\frac{d}{du}\left(h\frac{da_x^{(0)}}{du} \right)
+{\omega^2} a_x^{(0)}- \frac{2 h V_0}{u^2} a_x^{(0)}=0, \\
\label{eq:a_t1}
& h\frac{d^2}{du^2}a_t^{(1)}-  ( q^2a_t^{(1)}+iq\omega a_x^{(1)}) 
+ i  \omega h \frac{d}{du } a_u^{(1)} 
-\frac{2 V_0}{u^2} a_t^{(1)}=0, \\
\label{eq:a_x1}
& h\frac{d}{du}\left(h\frac{da_x^{(1)}}{du} \right)+ 
\omega^2 a_x^{(1)}-iq\omega a_t^{(1)}
- q h \frac{d}{du} (h  a_u^{(1)})
-\frac{2 h V_0}{u^2} a_x^{(1)} = \frac{2 h \delta V}{u^2} a_x^{(0)} \,, \\
\label{eq:a_u1}
& 
- i \omega \frac{d}{du} a_t^{(1)}  + q h \frac{d}{du} a_x^{(1)} 
+ \omega^2 a_u^{(1)} - q^2 h \, a_u^{(1)} - \frac{2  h V_0}{u^2} a_u^{(1)}   = 0 \,, \\
\label{eq:a_t2}
& h\frac{\p^2}{\p u^2}a_t^{(2)}
+    \frac{\p^2}{\p x^2}a_t^{(2)}+i\omega \frac{\p}{\p x}a_x^{(2)} 
+ i \omega h \frac{\p}{\p u} a_u^{(2)}
- \frac{2 V_0}{u^2} a_t^{(2)}=\frac{2 \delta V}{u^2} a_t^{(1)}\sin qx\cos qx \,, \\
\label{eq:a_x2}
& h\frac{\p}{\p u}\left(h\frac{\p a_x^{(2)}}{\p u} \right)
+ \omega^2 a_x^{(2)}-i\omega \frac{\p}{\p x}a_t^{(2)} 
- h \frac{\p}{\p u} \left(h\frac{\p a_u^{(2)}}{\p x} \right)
-\frac{2h V_0 }{u^2}a_x^{(2)}=\frac{2 h \delta V \cos^2 qx }{u^2} a_x^{(1)} \,, \\
& - i \omega \frac{\partial}{\partial u} a_t^{(2)} - h \frac{\partial^2}{\partial u \partial x} a_x^{(2)}  + h \frac{\partial^2}{\partial x^2} a_u^{(2)} + \omega^2 a_u^{(2)} - \frac{2  h V_0 }{u^2} a_u^{(2)} = \frac{2  h \delta V}{u^2} a_u^{(1)} \sin q x \cos q x \,.
\end{align}

In order to simplify, we define the average physical quantities over the spatial 
direction $x$ on the range $2 \pi/q$ as  
\begin{align}
\label{average}
\overline{A}(u):=\frac{q}{2 \pi}\int^{2 \pi/q}_0 A(u,x)dx.  
\end{align}
Then, due to the periodicity of the perturbation along the $x$ direction, 
(\ref{eq:a_x2}) becomes a simple differential equation as 
\begin{align}
\label{eq:a_x2:average}
h\frac{d}{du}\left(h\frac{d \overline{a_x^{(2)}}}{du} \right)
+{\omega^2}\overline{a_x^{(2)}}
-\frac{2h V_0}{u^2}\overline{a_x^{(2)}}=\frac{h \delta V}{u^2} a_x^{(1)} \,,
\end{align}
namely, $\overline{a_x^{(2)}}$ decouples from $\overline{a_t^{(2)}}$, $\overline{a_u^{(2)}}$.

Furthermore in order to impose ingoing boundary condition at the horizon, we 
re-define the fields as 
\begin{align}
\label{cond:infalling}
a_i^{(n)}=e^{i\omega u_*}\xi_i^{(n)}, \qquad 
n=0,1,2,\cdots
\end{align}
for $i = (t,x,u)$, where
\bea
u_* \equiv \int^u \frac{du}{ h(u) } \,.
\eea
Then, 
from (\ref{eq:a_x0}), (\ref{eq:a_t1}), (\ref{eq:a_x1}), (\ref{eq:a_x2:average}), we obtain 
\begin{align}
\label{eq:xi_x0}
& h\frac{d^2}{du^2}\xi_x^{(0)}
+\left(h'  + {2i\omega} \right)\frac{d\xi_x^{(0)}}{du}-\frac{2 V_0}{u^2} \xi_x^{(0)}=0 \,,  \\
\label{eq:xi_t1}
& h\frac{d^2}{du^2}\xi_t^{(1)}+{2i\omega}\frac{d\xi_t^{(1)}}{du}
+\left[\frac{- i \omega h'}{ h}-\frac{\omega^2}{ h}- {q^2}-\frac{2 V_0}{u^2}  \right]
\xi_t^{(1)} \nonumber \\
& \quad \quad \quad \quad \quad  ={iq\omega}\xi^{(1)}_x 
+ {\omega^2} \xi_u^{(1)} - i \omega h  \frac{d\xi_u^{(1)}}{du}   \,, \\
\label{eq:xi_x1}
& h \frac{d^2}{du^2}\xi_x^{(1)}
+ \left(h' + {2i\omega} \right)\frac{d\xi_x^{(1)}}{du}- \frac{2 V_0}{u^2}  \xi_x^{(1)} 
\nonumber \\
& \quad \quad \quad  \quad \quad  =  \left( {i q \omega} + q h' \right) \xi_u^{(1)}  
+ q h \frac{d \xi_u^{(1)} }{du}  
 + \frac{i q \omega}{ h}\xi_t^{(1)}+\frac{2 \delta V}{u^2} \xi_x^{(0)}, \\
\label{eq:xi_u1}
& \frac{\omega^2}{ h}  \xi_t^{(1)} - i \omega \frac{d \xi_t^{(1)}}{du} + {i q \omega}  \xi_x^{(1)} + q h  \frac{d \xi_x^{(1)}}{du} + \left( \omega^2 - q^2 h - \frac{2  h V_0}{u^2}  \right)   \xi_u^{(1)} = 0 \,,
\\
\label{eq:xi_x2:average}
& h\frac{d^2 \overline{\xi_x^{(2)}}}{du^2}
+\left( h' + {2i\omega} \right)\frac{d\overline{\xi_x^{(2)}}}{du}
-\frac{2 V_0}{u^2}  \overline{\xi_x^{(2)}}=\frac{\delta V}{u^2}\xi_x^{(1)} \,,
\end{align}
where $'$ means the $u$-derivatives. 
The zeroth order $\xi_i^{(0)}$ gives the conductivity without the lattice effects. 
By solving (\ref{eq:xi_x0}), we can obtain the zeroth order conductivity, {\it i.e.,} conductivity without the lattice. 

Let us first concentrate on the leading perturbation $\xi_i^{(1)}$. 
We are interested in the zero frequency delta function peak.  
For that purpose, it is enough to study the behavior of these equations 
at low frequency limit.   For that purpose, 
we expand those fields in small $\omega$ as 
\bea
\label{xiiomegaexpansion}
\xi_i^{(1)} = \xi_i^{(1),0} + \omega \xi_i^{(1),1}  + \omega^2 \xi_i^{(1),2} + \cdots  
\eea
for $i =t, x, u$. 

Then at the $O(\omega^0)$ order corresponding to the static limit, we have 
\bea
\label{eq:xi_t1omegazeroth}
&& h\frac{d^2}{du^2}\xi_t^{(1),0}
+\left(
- {q^2} -\frac{2 V_0}{u^2} \right)
\xi_t^{(1),0} = 0  \,,
\, \\
\label{eq:xi_x1omegazeroth}
&& h \frac{d^2}{du^2}\xi_x^{(1),0}
+ h' \frac{d\xi_x^{(1),0}}{du}- \frac{2 V_0}{u^2}  \xi_x^{(1),0} 
 =   q h'   \xi_u^{(1),0}  
+ q h \frac{d \xi_u^{(1),0}}{du}   
+\frac{2 \delta V}{u^2} \xi_x^{(0),0}, \\
\label{eq:xi_u1omegazeroth}
&& 
 q h  \frac{d \xi_x^{(1),0}}{du} + \left( - q^2 h - \frac{2  h V_0}{u^2}  \right)   \xi_u^{(1),0} = 0 \,. 
\eea
Therefore in this static limit $\omega \to 0$, $\xi_t^{(1),0}$ decouples from $\xi_x^{(1),0}$ 
and $\xi_u^{(1),0}$. Then it is determined by solving the equation (\ref{eq:xi_t1omegazeroth}), 
which gives unique radial coordinate $u$ dependent solution once we give the following two boundary condition; 
The non-normalizable mode of $\xi_t^{(1),0}$ must vanish at the boundary. $\xi_t^{(1),0}$ must vanish at the horizon, so that Wilson loop 
\bea
e^{\int A_\mu dx^\mu} \sim e^{\int A_\tau d\tau}
\eea
vanishes on the trivial cycle at the horizon, in the Euclid signature. 
These boundary condition determines that 
\bea
\label{xitzerofrequency}
\xi_t^{(1),0} = 0 \,.
\eea

Given this, for the next order $O(\omega^1)$, we have 
\bea
\label{eq:xi_t1omega1th}
&&  h\frac{d^2}{du^2}\xi_t^{(1),1}
+  \left( - {q^2}-\frac{2 V_0}{u^2}  \right)
\xi_t^{(1),1}
= {iq} \xi^{(1),0}_x  
- i h  \frac{d\xi_u^{(1),0}}{du}
\,, \quad \quad \\
\label{eq:xi_x1omega1th}
&& h \frac{d^2}{du^2}\xi_x^{(1),1}
+h' \frac{d\xi_x^{(1),1}}{du}
+ {2i} \frac{d\xi_x^{(1),0}}{du}
- \frac{2 V_0}{u^2}  \xi_x^{(1),1} 
\nonumber \\
&& \quad \quad \quad  \quad \quad  
=  {i q } \xi_u^{(1),0}   
+ q h'  \xi_u^{(1),1}  
+ q h \frac{d \xi_u^{(1),1} }{du}    
  +\frac{2 \delta V}{u^2} \xi_x^{(0),1}, \\
\label{eq:xi_u1omega1th}
&&  
 {i q } \xi_x^{(1),0} 
+ q h  \frac{d \xi_x^{(1),1}}{du} 
+ \left(  - q^2 h - \frac{2  h V_0}{u^2}  \right)   \xi_u^{(1),1} = 0 \,.
\eea
From (\ref{eq:xi_u1omegazeroth}) and (\ref{eq:xi_u1omega1th}), we can write down $\xi_u^{(1),0}$, $\xi_u^{(1),1}$ in terms of $\xi_x^{(1),0}$, $\xi_x^{(1),1}$ as 
\bea
\xi_u^{(1),0} &=& \frac{q}{ q^2  + \frac{2   V_0}{u^2} }   \frac{d \xi_x^{(1),0}}{du} \,, \\
\xi_u^{(1),1} &=& \frac{q}{ q^2  + \frac{2   V_0}{u^2} }  \left( \frac{d \xi_x^{(1),1}}{du}  
+ 
 \frac{i }{ h}  
  \xi_x^{(1),0}
  \right)   \,.
\eea
These give
\bea
\label{xiuintermsofxix}
\xi_u^{(1)}  
&=& \frac{q}{ q^2  + \frac{2   V_0}{u^2} } \left(  \frac{d \xi_x^{(1)}}{du} 
+  \frac{i  \omega}{ h} 
 \xi_x^{(1)}
\right) + O(\omega^2)  \,.
\eea
From this, it is straightforward to check that at the horizon where $h\to 0$, $\xi_\mu \xi^\mu$ is divergent-free.
\footnote{
To see this, note that by using the regularity of $\xi_t$ at the horizon, (\ref{eq:xi_t1omega1th}) gives
\bea
\xi_t^{(1),1} = -  \frac{i q}{ q^2  + \frac{2   V_0}{u^2} } \, \xi_x^{(1),0} \,.
\eea
So we have 
\bea
\xi_t^{(1)} =   - \omega \frac{i q}{ q^2  + \frac{2   V_0}{u^2} } \, \xi_x^{(1),0} + O(\omega^2) \,. 
\eea
Therefore, we have, at the horizon where $h \to 0$, neglecting $O(\omega^3)$, 
\bea
\xi_\mu \xi_\nu g^{\mu\nu} &=& (\xi_t)^2 g^{tt} + (\xi_x)^2 g^{xx} + (\xi_u)^2 g^{uu} \nonumber\\
&=& \frac{u^2}{L^2 } h^{-1} \omega^2 \left(\frac{ q}{ q^2  + \frac{2   V_0}{u^2} }\right)^2  (\xi_x^{(1),0})^2 
+ \frac{u^2}{L^2 } (\xi_x^{(1),0} + \omega \xi_x^{(1),1}+ \omega^2 \xi_x^{(1),2} )^2 \nonumber \\
&& +  \frac{u^2}{L^2} h \left( \frac{q}{ q^2  + \frac{2   V_0}{u^2} } \left(  \frac{d \xi_x^{(1),0}}{du} 
+ \omega   \left( \frac{d \xi_x^{(1),1}}{du}  
+  \frac{i }{ h} 
 \xi_x^{(1),0}
\right)  \right)  + \xi_u^{(1),2}\right)^2 + O(\omega^3) \,. \quad \quad \quad 
\eea
Here, by using the similar argument, from (\ref{eq:xi_u1}), we can see $\xi_u^{(1),2}$ diverges 
at the horizon as
\bea
\xi_u^{(1),2} = O(h^{-1}) \,.
\eea
Let's consider the leading divergent terms in above $\xi_\mu \xi_\nu g^{\mu\nu}$. 
Under the regularity condition for $\xi_x$, the leading divergent terms which blow up as $O(h^{-1})$ are  
\bea
\xi_\mu \xi_\nu g^{\mu\nu} &=&   \frac{u^2}{L^2 } h^{-1} \omega^2 \left(\frac{ q}{ q^2  + \frac{2   V_0}{u^2} }\right)^2  (\xi_x^{(1),0})^2 
 +  \frac{u^2}{L^2}\omega^2 h \left( \frac{q}{ q^2  + \frac{2   V_0}{u^2} }   
  \frac{i }{ h}  \xi_x^{(1),0} \right)^2 + O(h^0) \,. \quad \quad  \quad 
\eea
So at least up to $O(\omega^2)$, the leading divergent terms, which behave as $h^{-1}$ in $\xi_\mu \xi_\nu g^{\mu\nu}$,  
cancel. 
}

By plugging this (\ref{xiuintermsofxix}) back to the equations (\ref{eq:xi_t1}) and (\ref{eq:xi_x1}), we obtain differential equations for $\xi_t$ and $\xi_x$ up to order $O(\omega^2)$ accuracy as,
\bea
\label{finalsmallomegaeqone}
&& h\frac{d^2}{du^2}\xi_t^{(1)}  
+\left[  
-{q^2}-\frac{2 V_0}{u^2}  \right]
\xi_t^{(1)}={iq\omega}\xi^{(1)}_x 
- i \omega h  \frac{d }{du}  \left(  \frac{q}{ q^2  + \frac{2 V_0}{u^2} }  \frac{d \xi_x^{(1)}}{du}  \right) 
+ O(\omega^2) \,, \nonumber  \\
\\
\label{finalsmallomegaeqtwo}
&& h \frac{d^2}{du^2}\xi_x^{(1)}
+ \left(h' +{2i\omega} \right)\frac{d\xi_x^{(1)}}{du}- \frac{2 V_0}{u^2}  \xi_x^{(1)} 
\nonumber \\
&& \quad \quad \quad   = {\frac{q  \left( {i q \omega} + q h' \right) }{ q^2  + \frac{2 V_0}{u^2} } \left(  \frac{d \xi_x^{(1)}}{du} 
+  \frac{i  \omega}{ h}  \xi_x^{(1)} \right) }
+ q h \frac{d }{du}  \left(
\frac{q}{ q^2  + \frac{2  V_0}{u^2} } \left(  \frac{d \xi_x^{(1)}}{du} 
+  \frac{i  \omega}{ h} 
 \xi_x^{(1)}
\right) \right) \nonumber \\
&& \quad \quad \quad \quad  
 +\frac{2 \delta V}{u^2} \xi_x^{(0)} + O(\omega^2), 
\eea 
Note that terms like $\omega \xi_t^{(1)}$ are $O(\omega^2)$. 
We can solve those equations and resultantly we can determine the conductivity 
in the low frequency limit. The results allow us to check if 
zero frequency delta function peak exists or not at this order. 
 
Conductivity is given by
\bea
\sigma \equiv \frac{A_{x,u}}{ F_{xt}} = \frac{a_{x,u}}{ a_{t,x} +  i \omega a_x } \,,
\eea
which is generically position dependent. 
Here in numerator, dominant term is a normalizable mode, and in the denominator, dominant term is non-normalizable mode.  
Expanding $A_x$ and $A_t$ by $\epsilon$, 
we obtain, 
\bea
\sigma = \sigma^{(0)}(\omega) + \epsilon \cos q x \, \sigma^{(1)}(\omega) + \epsilon^2 \sigma^{(2)}(\omega, x) + \cdots
\eea
and 
\bea
\label{sigmapertubativezero}
 \sigma^{(0)}(\omega) &=& {1} + \frac{ \xi_x^{'(0)}(0)}{i \omega \xi_x^{(0)}(0)} \,, \\
\label{sigmapertubativeone}
\sigma^{(1)}(\omega) &=& 
-\frac{i \xi_x^{'(1)}(0)}{\omega   \xi_x^{(0)}(0)}
+\frac{i \xi_x^{(1)}(0)   \xi_x^{'(0)}(0)}{\omega \xi_x^{(0)}(0)^2}
+\frac{i q \xi_t^{(1)}(0)}{ \omega  \xi_x^{(0)}(0)}
+\frac{q \xi_t^{(1)}(0)   \xi_x^{'(0)}(0)}{\omega^2 \xi_x^{(0)}(0)^2} \,, \quad 
\eea
where $'$ is $u$-derivative.
If we look at the spatially averaged part of the conductivity $\overline{\sigma}$ as 
\bea
\label{def:averaged_cond}
 \overline{\sigma(\omega)}  
&\equiv& \frac{q}{2 \pi} \int_0^{2 \pi/q} \sigma dx \nonumber \\
&\equiv& \overline{\sigma}^{(0)}+\epsilon \overline{\sigma}(\omega)^{(1)}
+\epsilon^2 \overline{\sigma}(\omega)^{(2)}+\cdots \,,
\eea
then we obtain, 
\bea 
\label{averaged_cond}
 \overline{\sigma}^{(1)}(\omega) &=& 0, \nonumber \\
 \overline{\sigma}^{(2)}(\omega) &=& 
\frac{\overline{{\xi'}_x^{(2)}}(0)}{i\omega \xi_x^{(0)}(0)}
-\frac{\overline{\xi_x^{(2)}}(0) {\xi'}_x^{(0)}(0) }{i\omega \xi_x^{(0)}(0)^2}  
+ \frac{q \xi_t^{(1)}(0) \xi_x^{(1)}(0) }{2 i \omega  \xi^{(0)}(0)^2 } 
+ \frac{ \xi_x^{(1)}(0)^2 \xi_x^{(0)'}(0) }{2 i \omega \xi^{(0)}(0)^3 } \nonumber \\
&&
- \frac{\xi_x^{(1)}(0) \xi_x^{(1)'}(0) }{2 i \omega \xi_x^{(0)}(0)^2} 
- \frac{q^2 \xi_t^{(1)}(0)^2 }{2  \omega^2  \xi_x^{(0)}(0)^2 } 
- \frac{q \xi_t^{(1)}(0) \xi_x^{(1)}(0) \xi_t^{(0)'}(0)  }{ \omega^2 \xi_x^{(0)}(0)^3 } \nonumber \\
&& 
+ \frac{q \xi_t^{(1)}(0)  \xi_x^{(1)'}(0) }{2 \omega^2 \xi_x^{(0)}(0)^2} 
- \frac{q^2 \xi_t^{(1)}(0)^2 \xi_x^{(0)'}(0)}{2 i \omega^3 \xi_x^{(0)}(0)^3} \,.
\eea
We would like to evaluate these perturbative corrections to the conductivity by the 
lattice effects. However, in the real-world experiments, we usually apply a homogeneous 
electric field and see the conductivity. 
This means, that we should choose the boundary condition such that 
inhomogeneous parts of the electric field are set to be zero. This corresponds to choosing 
non-normalizable modes of the $O(\epsilon)$ terms, $\xi_i^{(1)}(u)$ for $(i = t, x)$, are set to zero 
\bea
\label{homogeneityflux}
\xi_i^{(1)}(0) = 0 \quad \,, \quad (i = t, x) \,, 
\eea
since they have $\cos qx$ dependence. 
Therefore, above conductivity formula 
reduces to 
\bea
\label{finalpertconductivity}
\sigma^{(1)}(\omega) = 
-\frac{i \xi_x^{'(1)}(0)}{\omega   \xi_x^{(0)}(0)} \quad \,, 
\quad \overline{\sigma}^{(2)}(\omega) = 
\frac{\overline{{\xi'}_x^{(2)}}(0)}{i\omega \xi_x^{(0)}(0)}
-\frac{\overline{\xi_x^{(2)}}(0) {\xi'}_x^{(0)}(0) }{i\omega \xi_x^{(0)}(0)^2}  \,.
\eea
These are the quantities which we will evaluate. 

Without solving the equations of motion explicitly, we can guess how the solution behaves 
at the zero frequency limit, and therefore, how the zero frequency delta function peak behaves at $\omega \to 0$ limit 
at this stage. 

In the $\omega \to 0$ limit, as we showed in (\ref{xiiomegaexpansion}) and (\ref{xitzerofrequency}), 
we have 
\bea
\xi_t^{(1)} = O(\omega) \quad \,, \quad  \xi_x^{(0)} = O(1) \quad \,, \quad  \xi_x^{(1)} = O(1) \,.  
\eea
Similarly we can also confirm that 
\bea
\xi_x^{(2)} = O(1) \,. 
\eea
Then by plugging these into the perturbative results 
(\ref{finalpertconductivity}), 
we can obtain that 
\bea
\mbox{Im} \sigma^{(0)} \sim \frac{1}{\omega} \quad \,, \quad  \mbox{Im} \sigma^{(1)} \sim \frac{1}{\omega}  \quad \,, \quad  \mbox{Im} \overline{\sigma}^{(2)} \sim \frac{1}{\omega}  \,.
\eea
Therefore we can guess that imaginary of $\sigma$ has a simple pole structure $\sim \frac{1}{\omega}$, as far as we consider the lattice effect perturbatively. 
However this argument is very naive, since there is a possibility 
that the residue of  the pole  becomes zero 
and pole disappears. For example, in the normal phase without lattice structure, 
the above argument breaks down since $\xi_x^{(0)} $ becomes zero therefore 
the residue of the $\frac1\omega$ vanish.  
Therefore in order to confirm above expectation, we will now solve the 
equations of motion in more explicitly and obtain the solutions. Then,  
we would like to see  
if the imaginary of $\sigma$ has a pole as 
\bea 
\mbox{Im} \, \sigma \sim \frac1\omega  \,.
\eea
Once this behavior is confirmed, using the Kramers-Kronig relation 
\bea
\label{KKrelation}
\mbox{Im[$\sigma(\omega)$]} = - \frac1\pi {\cal{P}} \int_{-\infty}^{\infty} d \omega' \frac{\mbox{Re[$\sigma(\omega')$]}}{\omega' - \omega} \,,
\eea
we can conclude that 
there is a zero frequency delta function peak.  
However even if the delta function peak remains, 
how its residue (weight), $\omega \times \mbox{Im} \, \sigma $ changes as we vary $q$ is very 
nontrivial. 
We will see through the 
numerical analysis that the weight, which is $\omega \times \mbox{Im} \, \sigma$, decreases as we increase $q$. 

\subsection{Explicit examples for AC conductivities at low frequency limit}

In order to work on some explicit examples, 
let us consider examples where we take 
\bea
V_0 = d_1 u^2(1+d_2u^2)  \quad \,, \quad 
\delta V = u^2 \,.
\eea
Now we numerically calculate the conductivity $\sigma^{(0)}(\omega)$, 
$\sigma^{(1)}(\omega)$, $\bar \sigma^{(2)}(\omega)$, 
given by the formula~(\ref{sigmapertubativezero}) and (\ref{finalpertconductivity}) for various 
$q$ in the two typical cases, (i) $d_1=5$, $d_2=0$ and (ii) $d_1=5$, $d_2=2$, respectively. 
We have chosen the power for $V_0$ as $u^2$ and $u^4$. These choices are due to the 
fact that the asymptotic behaviors of the background charged scalar, which condensates,  
behaves as $\Psi^{background} \sim u$ or $\Psi^{background} \sim u^2$, see eq. (8) of \cite{Hartnoll:2008vx}.   

By imposing regularity condition at the horizon, 
\begin{align}
\xi^{(0)}_x(1)=\mbox{regular}, 
\end{align} 
we obtain the conductivity $\sigma^{(0)}(\omega)$ by numerically 
solving (\ref{eq:xi_x0}). Fig.~1 - 4 show $\sigma^{(0)}(\omega)$.  
As is seen from the figures for the Re[$\sigma^{(0)}(\omega)$], for the both cases, 
the gap appears.  
The energy gap in Fig.~4 is larger than the one in Fig.~2 because the 
``condensation'' $V_0(u) \sim |\Psi^{background}|^2$ in the case (ii) is larger 
than the one in the case (i).  
As we mentioned in the introduction, 
these results for conductivities are quite similar to the conductivity calculations for the 
holographic superconductor \cite{Hartnoll:2008vx}. Especially there are energy gaps and 
delta function peak at Re[$\sigma^{(0)} (\omega)$] $= 0$, 
which is seen from the imaginary parts of $\sigma_0$ behave 
as $\frac1\omega$.   
For in Fig.~2 and 4, both Re[$\sigma^{(0)}(\omega)$] approaches small but nonzero value at $\omega \to 0$, 
corresponding to the existence of the mass gap.  

By Kramers-Kronig relation (\ref{KKrelation}), 
the real part of the conductivity contains a delta function peak 
such as $\mbox{Re}[\sigma^{(0)}(\omega)]\simeq \pi C^{(0)}\delta(\omega)$ if the imaginary 
part of the conductivity contains a pole such as 
$\mbox{Im}[\sigma^{(0)}(\omega)]\simeq C^{(0)}/\omega$. Fig.~1 and 3 show that the pole exists 
at $\omega=0$ for both cases (i) and (ii). The best fitting curves determine the 
coefficients $C^{(0)}$ as $C^{(0)}=3.1$~(the case (i)) and $C^{(0)}=3.25$~(the case (ii)). 
 
\begin{figure}
 \begin{center}
   \includegraphics[width=60mm]{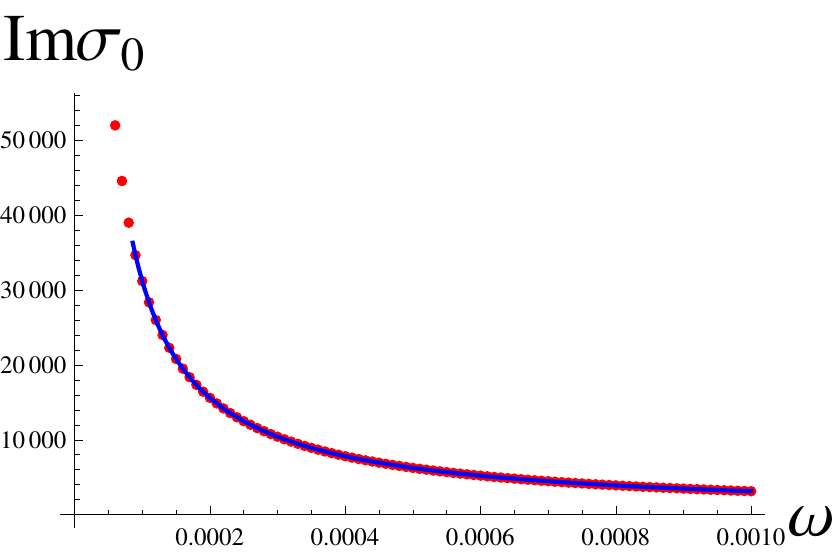}
  \caption{Im[$\sigma^{(0)}$] is plotted for various $\omega$~(red) in the case (i). 
The best fitting curve~(blue) is Im$[\sigma^{(0)}(\omega)]\simeq-9.4\times 10^{-4}+3.1/\omega$.}
 \end{center}
\end{figure}
\begin{figure}
 \begin{center}
  \includegraphics[width=60mm]{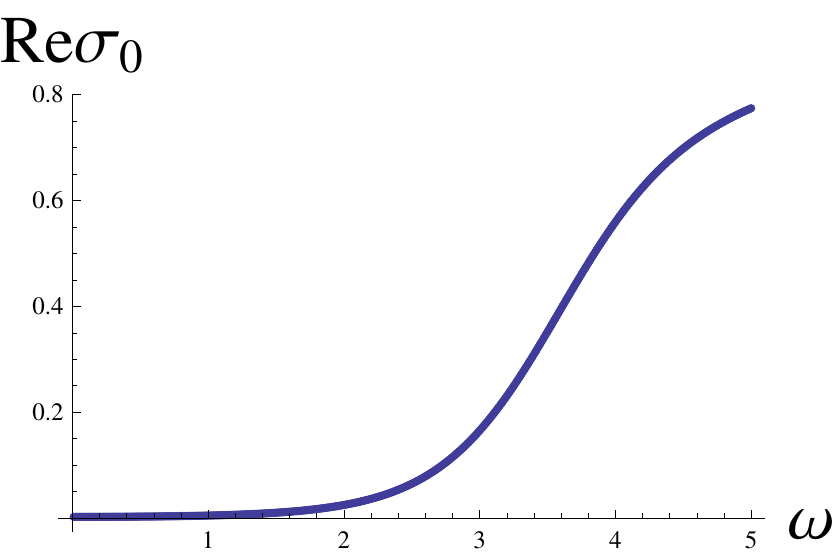}
\caption{Re[$\sigma^{(0)}$] is plotted for various $\omega$ in the case (i). 
There is a delta function peak at $\omega = 0$.}
 \end{center}
\end{figure}

\begin{figure}
 \begin{center}
  \includegraphics[width=60mm]{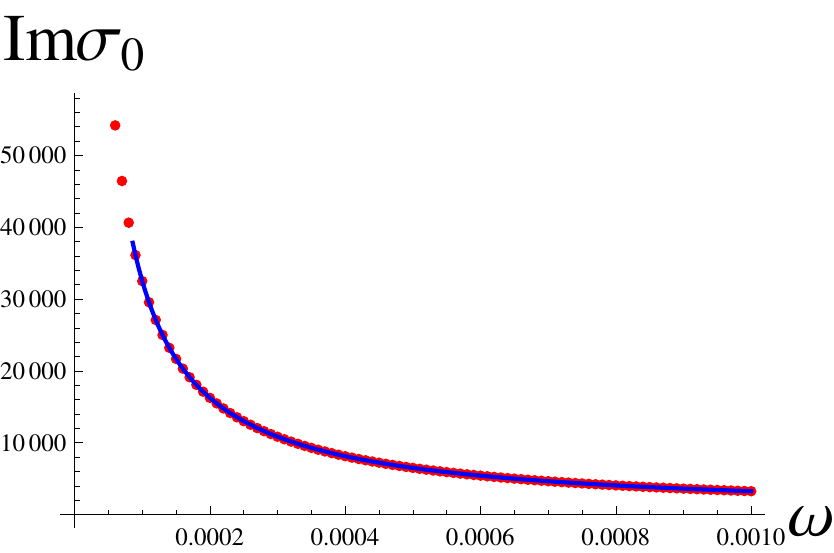}
 \caption{Im[$\sigma^{(0)}$] is plotted for various $\omega$~(red) in the case (ii). 
The best fitting curve~(blue) is Im$[\sigma^{(0)}(\omega)]\simeq-8.4\times 10^{-5}+3.25/\omega$.}
 \end{center}
\end{figure}
\begin{figure}
 \begin{center}
 \includegraphics[width=60mm]{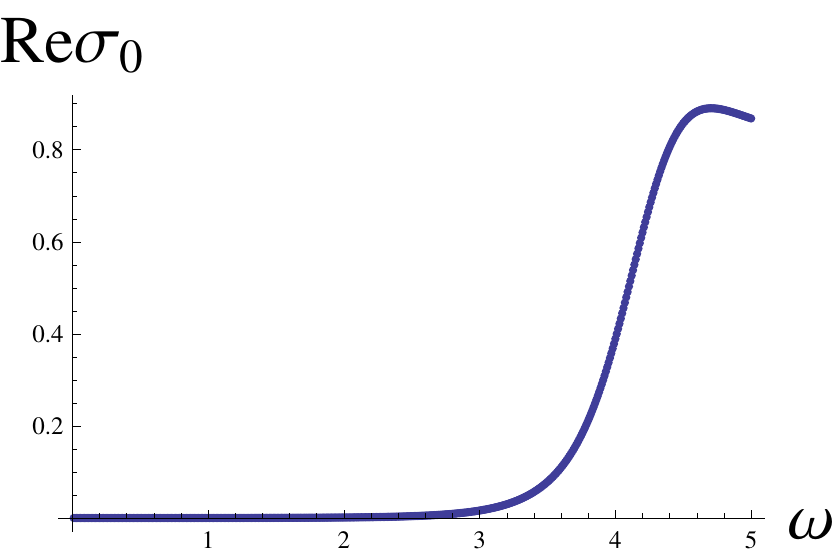}
  \caption{Re[$\sigma^{(0)}$] is plotted for various $\omega$ in the case (ii). 
  There is a delta function peak at $\omega = 0$.}
 \end{center}
\end{figure}

The first order conductivity $\sigma^{(1)}(\omega)$ is obtained by numerically 
solving the following equations from (\ref{finalsmallomegaeqone}) and (\ref{finalsmallomegaeqtwo});
\bea
&& h\frac{d^2}{du^2}\xi_t^{(1)}  
+\left[  
- {q^2} - 2 d_1(1+d_2u^2) \right]
\xi_t^{(1)} \nonumber \\ 
&& 
= {iq\omega}  \xi^{(1)}_x 
-   \frac{i \omega h   q}{ q^2  + {2 d_1(1+d_2u^2)} }  \frac{d^2 \xi_x^{(1)}}{du^2}  
+ \frac{4i \omega h q d_1d_2u}{(q^2+2  d_1(1+d_2u^2))^2}
\frac{d\xi_x^{(1)}}{du}
+ O(\omega^2) \,, \nonumber  \\
\eea
\bea
&& h \frac{d^2}{du^2}\xi_x^{(1)}
+ \left(h' + {2i\omega} \right)\frac{d\xi_x^{(1)}}{du}- {2 d_1(1+d_2u^2)}  \xi_x^{(1)} 
\nonumber \\
&& \quad 
= \left( \frac{q  \left( {i q \omega} + q h' \right) }{ q^2  + 2  d_1 (1 + d_2 u^2)}
- \frac{4 d_1 d_2 h q^2 u}{(q^2  + 2  d_1 (1 + d_2 u^2))^2}     \right) 
\, \left(  \frac{d \xi_x^{(1)}}{du} 
+ \frac{i  \omega}{ h}  \xi_x^{(1)} \right) 
\nonumber \\
&& \quad \quad
+ \frac{q^2 h }{ q^2  + {2   d_1 (1+d_2u^2) }  } \left(  \frac{d^2 \xi_x^{(1)} }{du^2} 
+  \frac{i  \omega}{ h} 
 \frac{d \xi_x^{(1)}}{du}
\right) 
- \frac{ i  \omega q^2 h' }{h \left( q^2  + {2   d_1 (1+d_2u^2) } \right) }    \xi_x^{(1)} \nonumber \\
&& \quad \quad  +2  \xi_x^{(0)} + O(\omega^2) \,, \quad \quad
\eea

To obtain the solution, we need to impose boundary conditions 
both at the horizon and the infinity. As we mentioned, in real-world experiments, 
we usually apply a homogeneous electric field and measure the conductivity. 
Therefore we shall impose a constant electric field condition at infinity. 
Since $O(\epsilon)$ part of the flux has $\cos q x$ dependence, we require that 
\begin{align}
\label{cond:electric}
E^{(1)}_x(0)= \left( i\omega a_x^{(1)}(0) + q a^{(1)}_{t}(0) \right) \cos q x = 0 \,,  
\end{align} 
for arbitrary $x$. 
We have chosen $\xi_x^{(1)}(0) = \xi^{(1)}_{t}(0) = 0$ as (\ref{homogeneityflux}), 
therefore (\ref{cond:electric}) is satisfied. 
We also require that the regularity condition for $\xi_x^{(1)}$ and $\xi_t^{(1)}$ at the horizon, which 
yield the ingoing condition for $a_x^{(1)}$ and $a_t^{(1)}$.  

Fig.~5 - 10 show Im[$\sigma^{(1)}(\omega)$] for various wavenumbers $q=2,\,5,\,10$ for each 
case. Let us define the coefficient $\tilde{C}(q)$ as 
$\mbox{Im}[\sigma^{(1)}(\omega)]=\tilde{C}(q)/\omega+O(1)$ 
near $\omega=0$. Then, the coefficient $\tilde{C}(q)$ can be read from the best fitting curves 
in Fig.~5 - 10 as $\tilde{C}(2)=0.41$,\,$\tilde{C}(5)=0.78$,\,and $\tilde{C}(10)=1.6$ 
for the case (i), while $\tilde{C}(2)=0.38$,\,$\tilde{C}(5)=0.73$,\,and 
$\tilde{C}(10)=1.58$ for the case (ii). 
This implies that the coefficient $\tilde{C}(q)$ increases as $q$ increases.  
\begin{figure}
 \begin{center}
  \includegraphics[width=6.3truecm,clip]{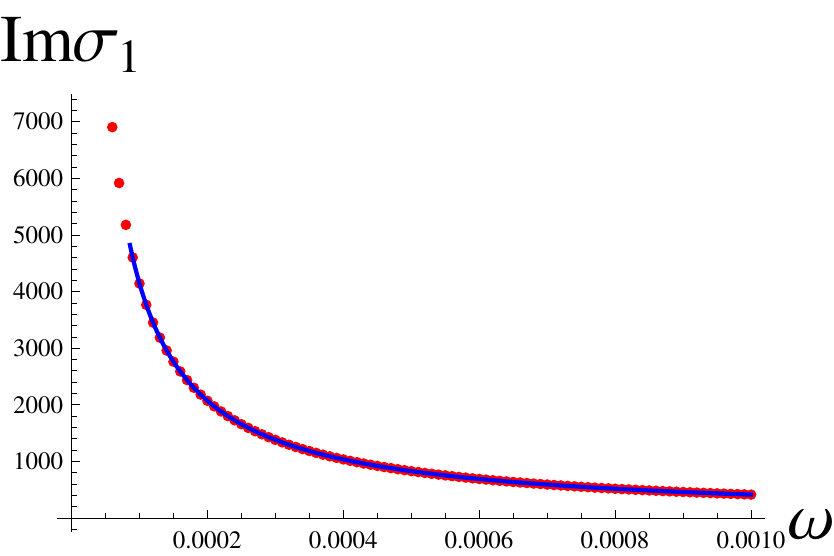}
  \caption{(color online) $\mbox{Im}[\sigma(\omega)^{(1)}]$ in the case (i) for 
$q=2$. The best fitting curve is 
$\sigma^{(1)}\simeq 9.2\times 10^{-5}+0.41/\omega$} 
 \end{center}
\end{figure}
\begin{figure}
 \begin{center}
  \includegraphics[width=6.3truecm,clip]{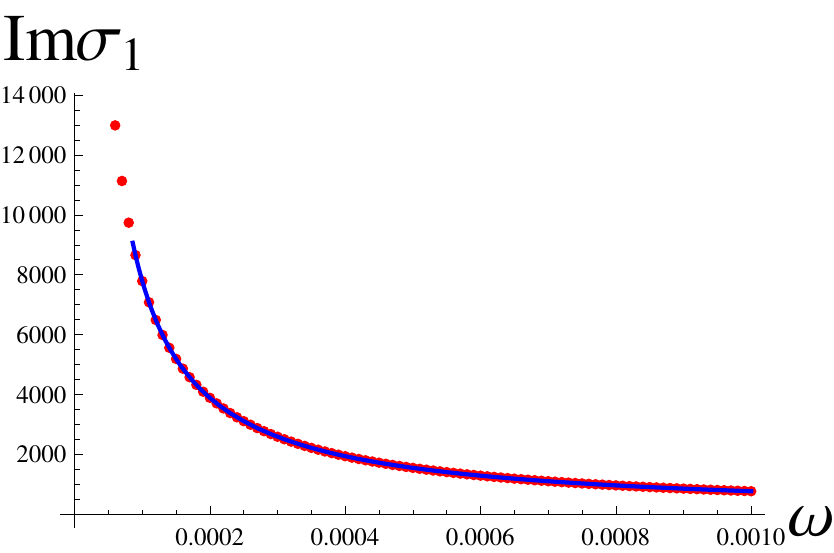}
  \caption{(color online) $\mbox{Im}[\sigma(\omega)^{(1)}]$ in the case (i) for  
$q=5$. The best fitting curve is 
$\sigma^{(1)}\simeq 0.0024+0.78/\omega$} 
 \end{center}
\end{figure}
\begin{figure}
 \begin{center}
  \includegraphics[width=6.3truecm,clip]{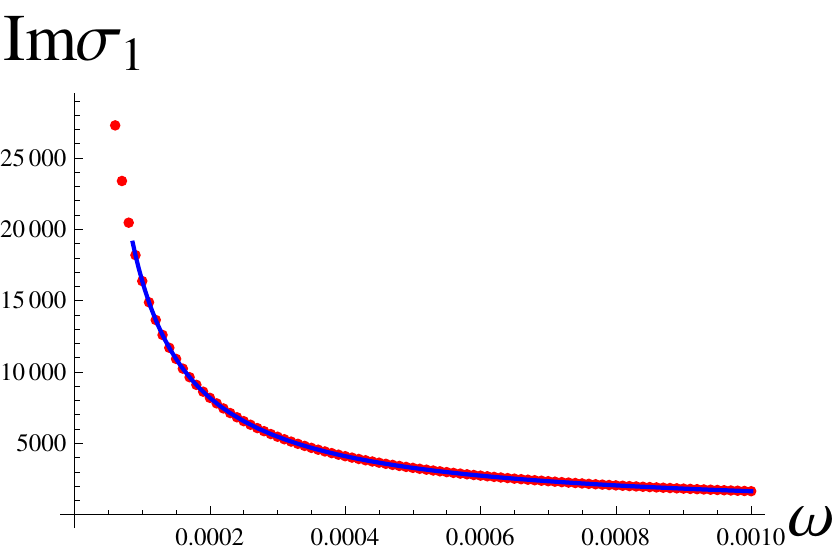}
  \caption{(color online) $\mbox{Im}[\sigma(\omega)^{(1)}]$ in the case (i) for  
$q=10$. The best fitting curve is 
$\sigma^{(1)}\simeq 0.72+1.6/\omega$} 
 \end{center}
\end{figure}
\begin{figure}
 \begin{center}
  \includegraphics[width=6.3truecm,clip]{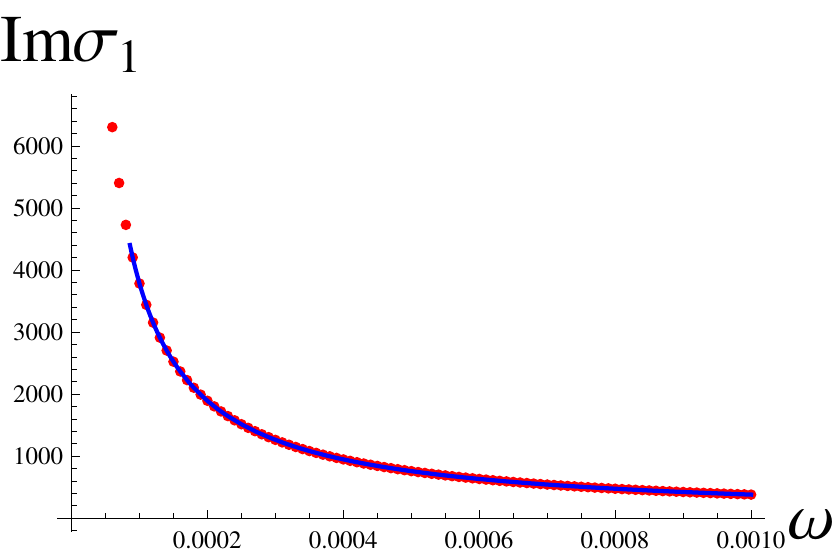}
  \caption{(color online) $\mbox{Im}[\sigma(\omega)^{(1)}]$ in the case (ii) for 
$q=2$. The best fitting curve is 
$\sigma^{(1)}\simeq 2.7\times 10^{-4}+0.38/\omega$} 
 \end{center}
\end{figure}
\begin{figure}
 \begin{center}
  \includegraphics[width=6.3truecm,clip]{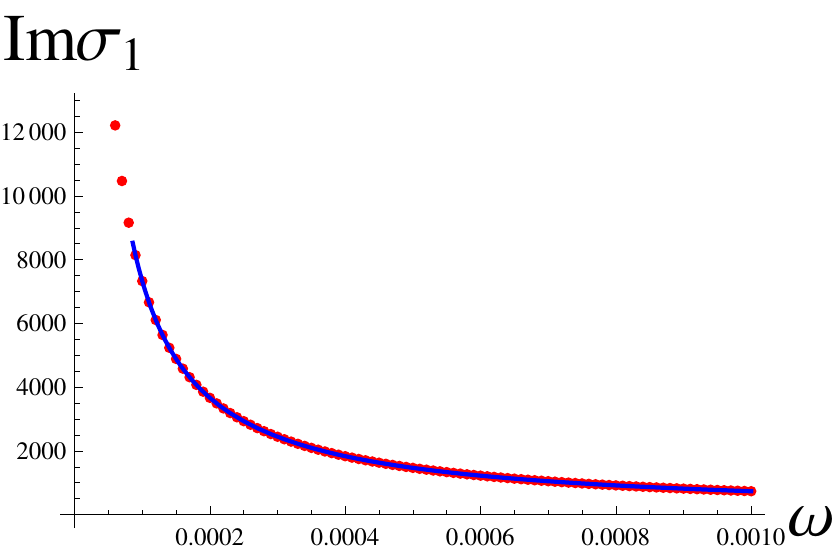}
  \caption{(color online) $\mbox{Im}[\sigma(\omega)^{(1)}]$ in the case (ii) for  
$q=5$. The best fitting curve is 
$\sigma^{(1)}\simeq 0.0024+0.73/\omega$} 
 \end{center}
\end{figure}
\begin{figure}
 \begin{center}
  \includegraphics[width=6.3truecm,clip]{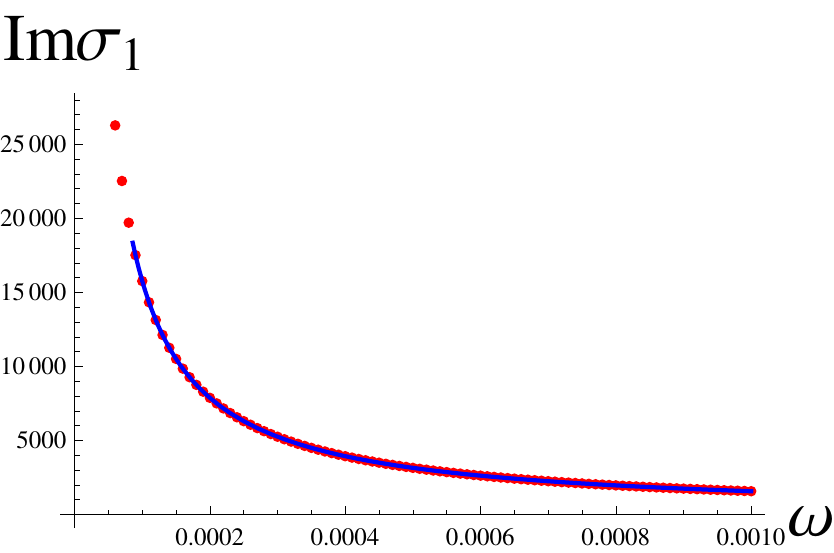}
  \caption{(color online) $\mbox{Im}[\sigma(\omega)^{(1)}]$ in the case (ii) for  
$q=10$. The best fitting curve is 
$\sigma^{(1)}\simeq 0.49+1.58/\omega$.} 
 \end{center}
\end{figure}

Finally, we show $q$-dependence of $\overline{\sigma}(\omega)^{(2)}$ in Figs.~11 - 16. 
$\overline{\sigma}(\omega)^{(2)}$ is numerically obtained by solving Eq. (\ref{eq:xi_x2:average}) under 
the regularity condition for $\bar \xi_x^{(2)}$ at the horizon and the constant electric field 
condition. 
Let us expand the imaginary part of the spatially averaged conductivity $\overline{\sigma}$
as 
\begin{align}
\mbox{Im}[\overline{\sigma}]=\frac{C(q)}{\omega}
=\frac{C^{(0)}+\epsilon C^{(1)}(q)+\epsilon^2 C^{(2)}(q)
+\cdots}{\omega}+O(1)
\end{align}
near the origin, $\omega=0$. Then, by definition, we immediately obtain $C^{(1)}(q)=0$.  
Therefore, the coefficient $C(q)$ is given by 
\begin{align}
C(q)=C^{(0)}+\epsilon^2 C^{(2)}(q).  
\end{align}

\begin{figure}
 \begin{center}
  \includegraphics[width=6.3truecm,clip]{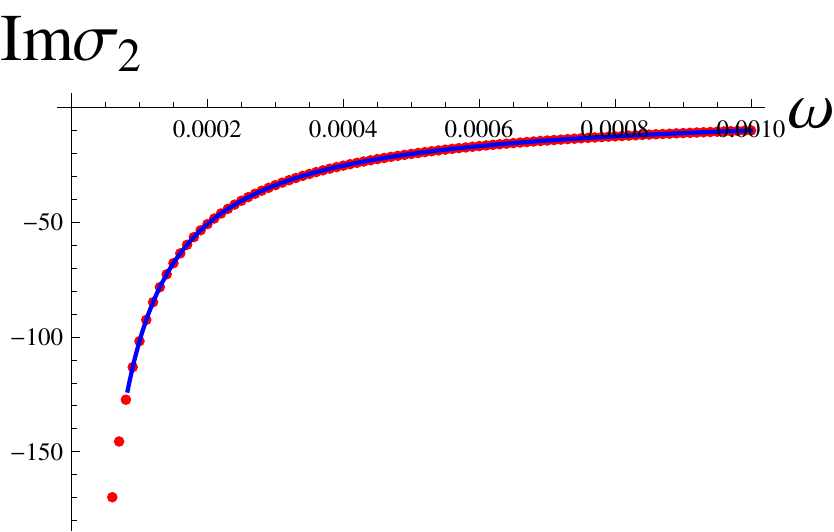}
  \caption{(color online) $\mbox{Im}[\sigma(\omega)^{(2)}]$ in the case (i) for 
$q=2$. The best fitting curve is 
$\sigma^{(2)}\simeq -1.3\times 10^{-6}-0.010/\omega$} 
 \end{center}
\end{figure}
\begin{figure}
 \begin{center}
  \includegraphics[width=6.3truecm,clip]{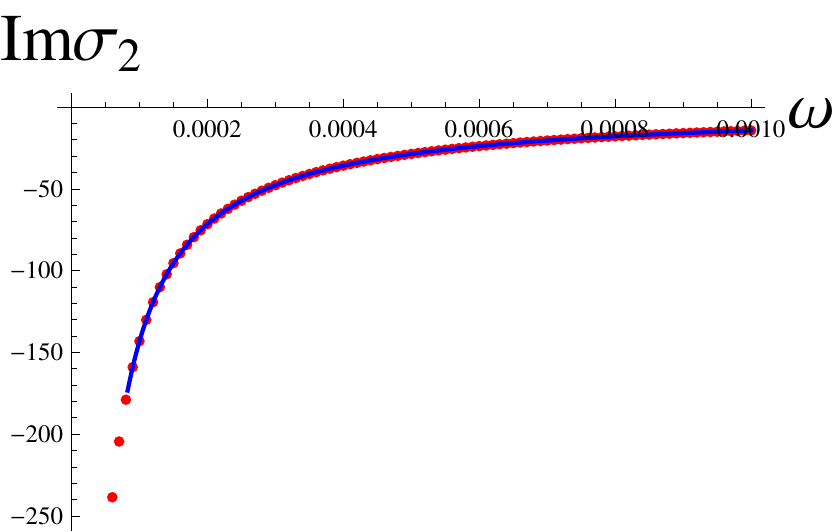}
  \caption{(color online) $\mbox{Im}[\sigma(\omega)^{(2)}]$ in the case (i) for  
$q=5$. The best fitting curve is 
$\sigma^{(2)}\simeq -2.5\times 10^{-6}-0.014/\omega$} 
 \end{center}
\end{figure}
\begin{figure}
 \begin{center}
  \includegraphics[width=6.3truecm,clip]{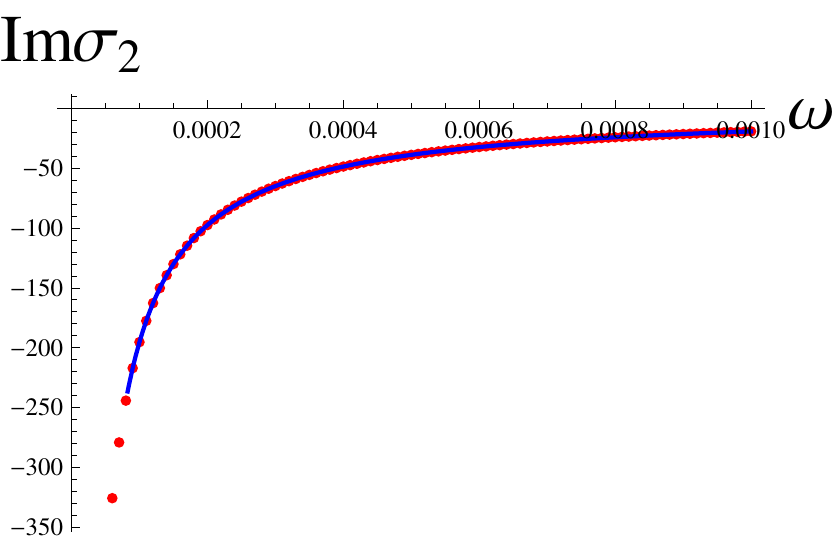}
  \caption{(color online) $\mbox{Im}[\sigma(\omega)^{(2)}]$ in the case (i) for  
$q=10$. The best fitting curve is 
$\sigma^{(2)}\simeq -3.9\times 10^{-6}-0.020/\omega$} 
 \end{center}
\end{figure}
\begin{figure}
 \begin{center}
  \includegraphics[width=6.3truecm,clip]{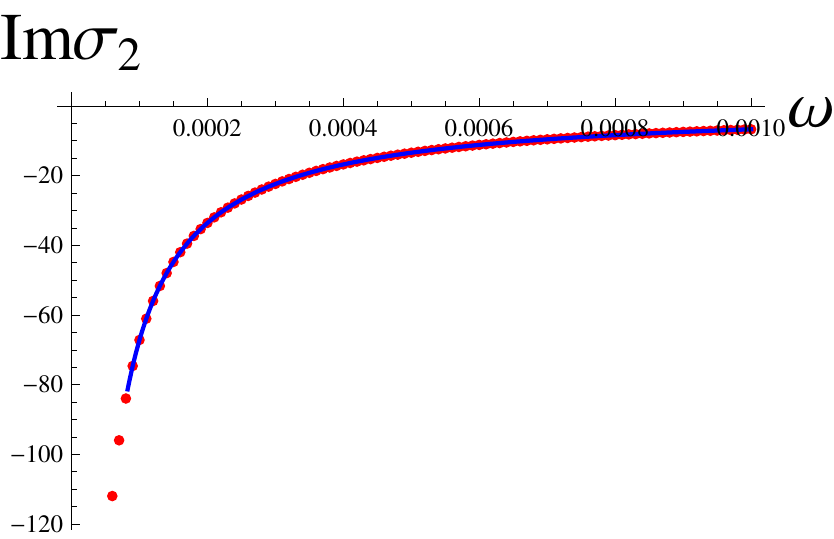}
  \caption{(color online) $\mbox{Im}[\sigma(\omega)^{(2)}]$ in the case (ii) for 
$q=2$. The best fitting curve is 
$\sigma^{(2)}\simeq -5.6\times 10^{-7}-0.0067/\omega$} 
 \end{center}
\end{figure}
\begin{figure}
 \begin{center}
  \includegraphics[width=6.3truecm,clip]{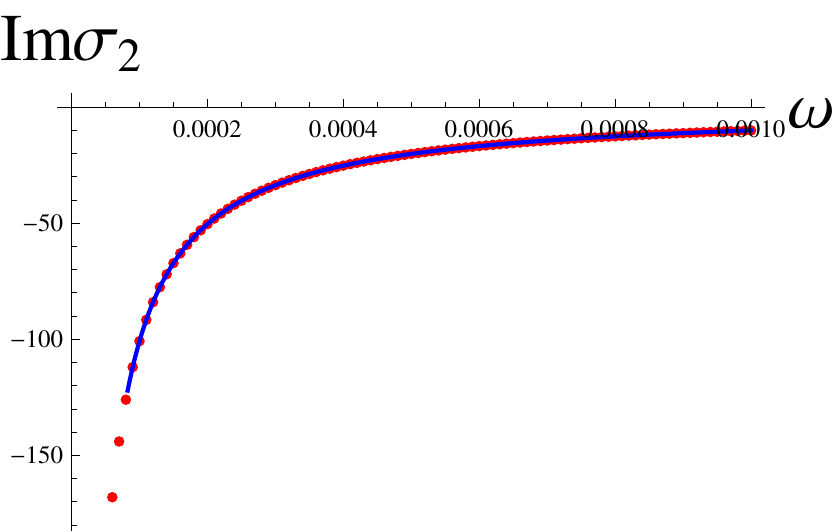}
  \caption{(color online) $\mbox{Im}[\sigma(\omega)^{(2)}]$ in the case (ii) for  
$q=5$. The best fitting curve is 
$\sigma^{(2)}\simeq 3.5\times 10^{-7}-0.010/\omega$} 
 \end{center}
\end{figure}
\begin{figure}
 \begin{center}
  \includegraphics[width=6.3truecm,clip]{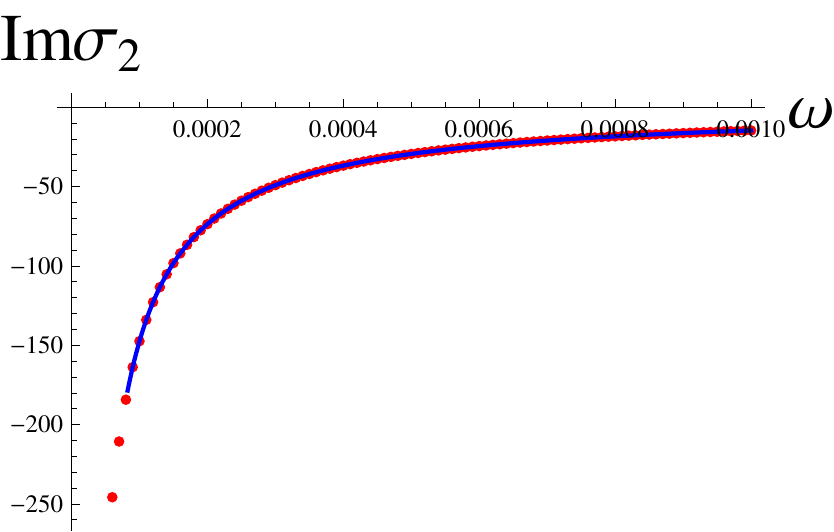}
  \caption{(color online) $\mbox{Im}[\sigma(\omega)^{(2)}]$ in the case (ii) for  
$q=10$. The best fitting curve is 
$\sigma^{(2)}\simeq 0.00015-0.015/\omega$.} 
 \end{center}
\end{figure}
The coefficient $C^{(2)}(q)$ can be read from the best fitting curves as 
in Fig.~11 - 16 as $C^{(2)}(2)=-0.010$,  $C^{(2)}(5)=-0.014$, and $C^{(2)}(10)=-0.020$ 
for the case (i), while $C^{(2)}(2)=-0.0067$, $C^{(2)}(5)=-0.010$, and 
$C^{(2)}(10)=-0.015$ for the case (ii). Therefore, the lattice effects reduce the coefficient 
$C(q)$ for any wavenumber $q$. Furthermore, we find that as $q$ increases, the coefficient 
$C^{(2)}(q)$, and therefore $C(q)$, decreases. 

By Kramers-Kronig relation, the real part of the conductivity contains a delta function peak 
if the imaginary part of the conductivity contains a pole. 
All these results suggest that, the magnitudes of the zero frequency delta function 
peak decrease by the 
lattice effects. This implies that in the holographic superconductor, the ``superfluid component'' 
of the conductivity decreases by the lattice effects.

\section{Conclusion and Discussion}
We studied the lattice effects on the toy model of holographic superconductor (superfluidity), 
massive $U(1)$ gauge boson model. Especially we studied how the 
zero frequency delta function peak on the real part of the conductivity is influenced by the lattice effects. 
Our analysis suggests that even though its weight reduces, 
the delta-function peak still remains even after the lattice effects are taken into account. 
This implies that  the superfluid component  remains with the lattice.  
We have seen also that, as the wavenumber of the lattice increases, the weight of the 
delta function peak decreases. 

However in order to get conclusive results, clearly we need to study things in more great detail. 
In our toy holographic superconductivity (superfluidity) model, we have neglected two important ingredients, the dynamics of the charged scalar field and also the gravity. 
For the charged scalar field, instead of treating it as a dynamical field, 
we have given its VEV by hand as an input. This nonzero VEV corresponds to the $U(1)$ 
symmetry breaking and yields the mass term for the gauge boson.  
Even though we vary this VEV and its radial profile   
through the several parameters of our system, we have seen that 
the zero frequency delta function peak remains. 
Therefore, we expect that  the results will not be modified much even after we have taken into 
account the charged scalar dynamics. 
However of course, 
 it is better to confirm this point in more explicitly by taking into account the dynamics of the 
charged scalar field \cite{IizukaMaedawork}. 

We have also neglected the effects of the gravity.  
There are two important effects associated with the gravity dynamics; 
The first one is the back reaction of the lattice effects to the geometry, since  
we have used the background geometry which does not possess the lattice effects. 
This correction can be calculated perturbatively, as is studied, for example, in \cite{Maeda:2011pk}. 
This induces the perturbative corrections to the background geometry, and 
how the perturbative correction appears on the geometry depends on 
how we introduce it. 
Suppose the lattice effects are $O(\epsilon)$, then, depending on the lattice effects which 
appear to the energy-momentum tensor at either  
$O(\epsilon)$ or $O(\epsilon^2)$, the order of back reaction to the  
geometry is different. 
If we introduce the perturbative lattice effects on the chemical potential as \cite{Maeda:2011pk} 
to the background with nonzero chemical potential background, then 
the back reaction of the lattice effects appear as $O(G_N \, \epsilon)$,  
then it cannot be neglected unless we take the probe limit. 
On the other hand, if we introduce the perturbative lattice effects by introducing 
neutral scalar field as \cite{Horowitz:2012ky}, then its back reaction appears as $O(G_N \,  \epsilon^2)$ therefore it can be neglected at the leading order in $\epsilon$ without taking the 
probe limit. 

There is also gravity effects on the conductivity calculations. As we quote in the introduction, the 
delta function peak does not appear in the normal phase, where $U(1)$ symmetry is preserved. 
This can be seen for example, by the fact that without gravity effect, on the normal phase, 
the equations of motion for the gauge boson $A_\mu$ 
admits only trivial constant solution\footnote{In fact if we re-write the equations of motion for the gauge boson $A_\mu$ as 
Schrodinger equation and treating the calculation for the 
conductivity as scattering problem \cite{Horowitz:2009ij}, the potential vanishes if we neglect the 
gravity effect on the normal phase. Therefore the conductivity is always unity, and we cannot see any 
delta function peak.}.  
Once we take into account the gravity, we can see the delta function peak, originated from the 
translational invariance of the system. In \cite{Horowitz:2012ky}, by taking into account the 
gravity effect, it is shown that the zero frequency delta function peak becomes flatten once we take into account the 
lattice effects. 
It is interesting to see how the results in this paper are influenced if we take into account the gravity effect.

Finally even if we take into account all of above effects, it is not clear if 
the perturbative analysis, as we have done in this paper, 
is enough to give us conclusive results. 
It is possible that there are non-perturbative corrections to the conductivity 
by the lattice effects, which 
significantly influence the delta function peak. In such cases, 
we may have to rely fully on the numerical analysis.  
We left these open questions for the future projects.

\acknowledgments
We would like to thank K. Hashimoto and G. Horowitz  
for discussions and comments on the draft. 
N.I. would like to thank Mathematical physics laboratory in RIKEN for very kind hospitality. 
N.I. is supported in part by the COFUND fellowship at CERN. 
K.M. is supported in part by 
MEXT/JSPS KAKENHI Grant Number 23740200.


\appendix

\section{A real part of the conductivity}
In this appendix, we summarize the numerical data for the real part of the conductivity. 
In Figs. 2 and 4, we show the Re[$\sigma(\omega)^{(0)}$] in the case of $d_1=5$, $d_2=0$ and $d_1=5$, $d_2=2$, respectively.  
It is clear that there is an energy gap approximately in the region $0\le \omega \le 3$ for both cases.

In Figs. 17 - 22, Re[$\sigma(\omega)^{(i)}$]~($i=1,2$) are plotted in the case of $d_1=5$, $d_2=0$ and $d_1=5$, 
$d_2=2$, respectively for each $q=2,5$. In any case, the real part approaches a constant in the limit $\omega\to 0$. 
\begin{figure}
 \begin{center}
  \includegraphics[width=6.3truecm,clip]{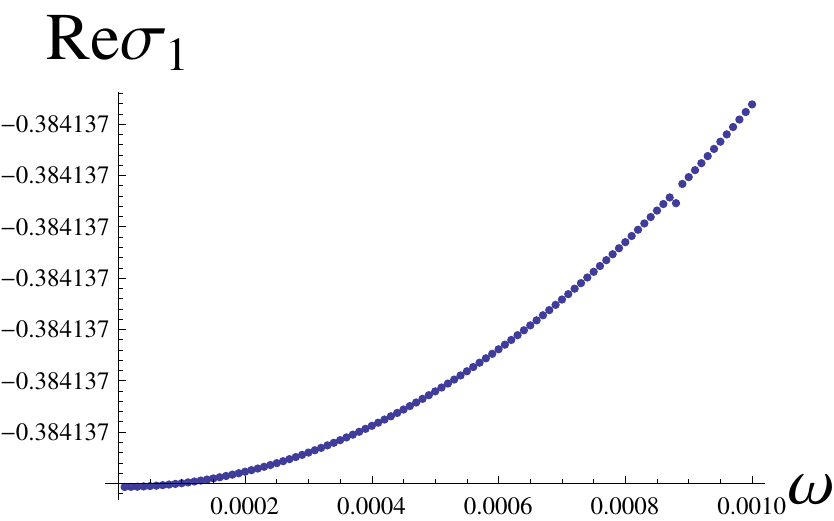}
  \caption{(color online) $\mbox{Re}[\overline{\sigma(\omega)^{(1)}}]$ in the case of 
$d_1=5,\,d_2=0,\,q=2$.} 
 \end{center}
\end{figure}
\begin{figure}
 \begin{center}
  \includegraphics[width=6.3truecm,clip]{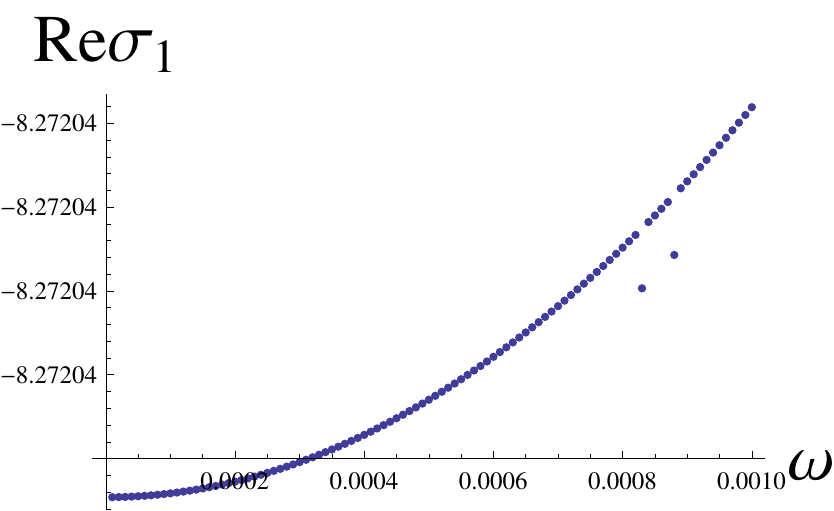}
  \caption{(color online) $\mbox{Re}[\overline{\sigma(\omega)^{(1)}}]$ in the case of 
$d_1=5,\,d_2=0,\,q=5$.} 
 \end{center}
\end{figure}
\begin{figure}
 \begin{center}
  \includegraphics[width=6.3truecm,clip]{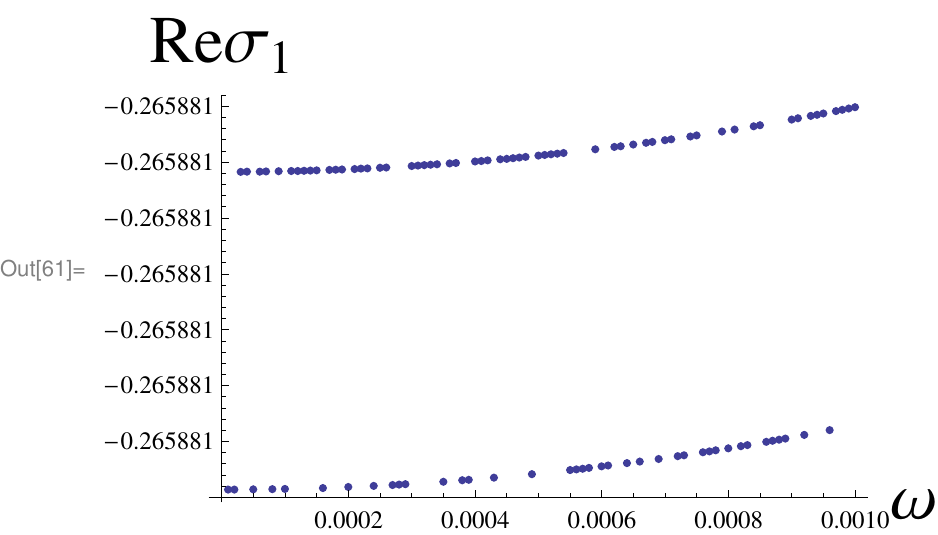}
  \caption{(color online) $\mbox{Re}[\overline{\sigma(\omega)^{(1)}}]$ in the case of 
$d_1=5,\,d_2=2,\,q=2$.} 
 \end{center}
\end{figure}
\begin{figure}
 \begin{center}
  \includegraphics[width=6.3truecm,clip]{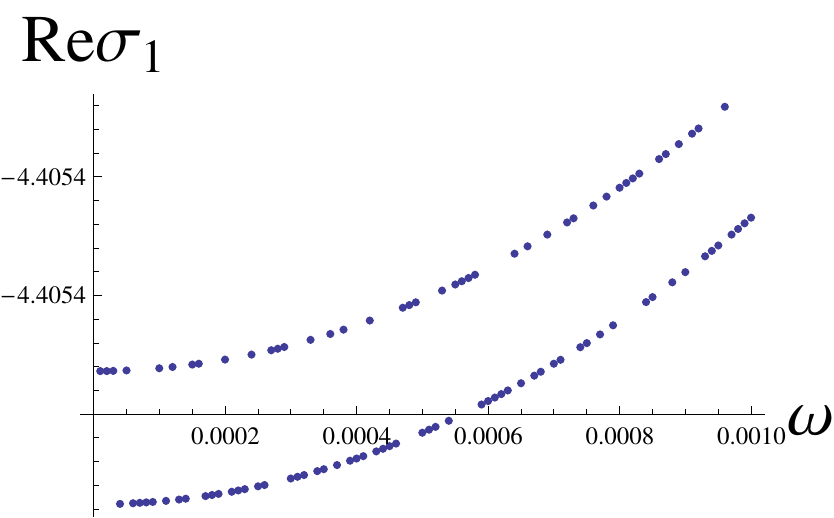}
  \caption{(color online) $\mbox{Re}[\overline{\sigma(\omega)^{(1)}}]$ in the case of 
$d_1=5,\,d_2=2,\,q=5$.} 
 \end{center}
\end{figure}
\begin{figure}
 \begin{center}
  \includegraphics[width=6.3truecm,clip]{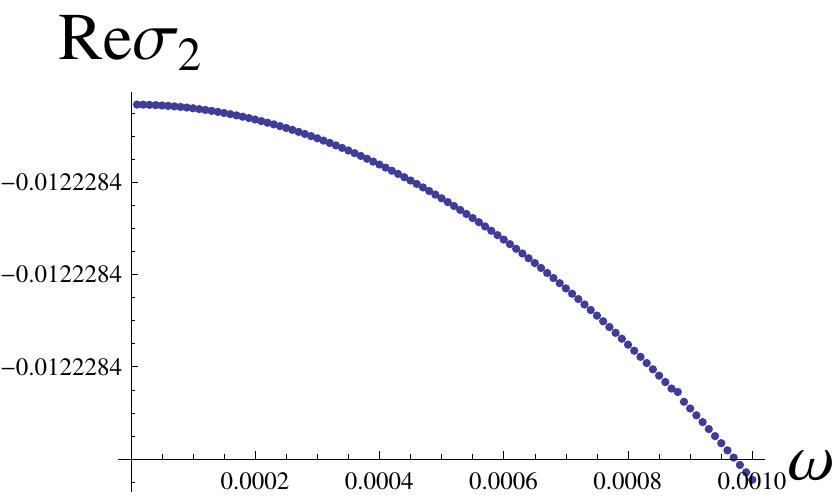}
  \caption{(color online) $\mbox{Re}[\overline{\sigma(\omega)^{(2)}}]$ in the case of 
$d_1=5,\,d_2=0,\,q=2$.} 
 \end{center}
\end{figure}
\begin{figure}
 \begin{center}
  \includegraphics[width=6.3truecm,clip]{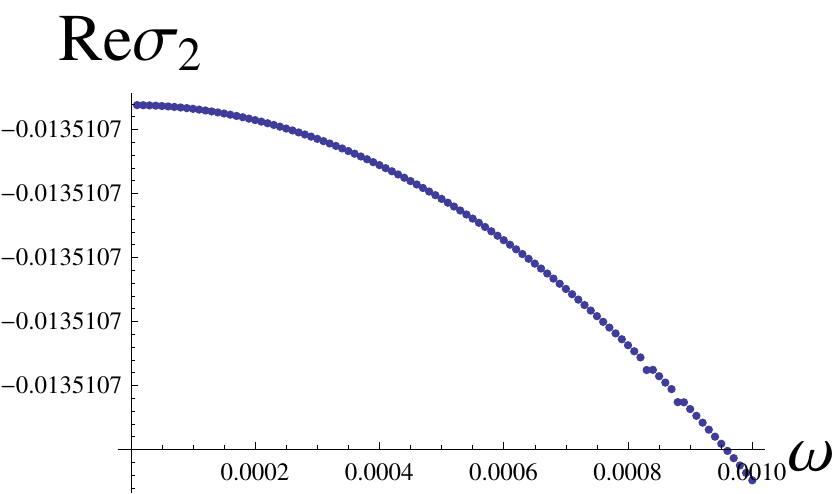}
  \caption{(color online) $\mbox{Re}[\overline{\sigma(\omega)^{(2)}}]$ in the case of 
$d_1=5,\,d_2=0,\,q=5$.} 
 \end{center}
\end{figure}
\begin{figure}
 \begin{center}
  \includegraphics[width=6.3truecm,clip]{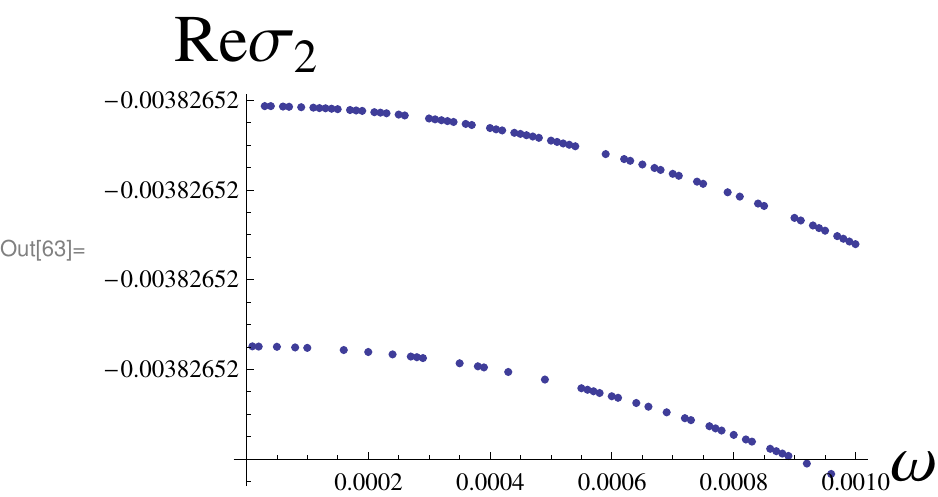}
  \caption{(color online) $\mbox{Re}[\overline{\sigma(\omega)^{(2)}}]$ in the case of 
$d_1=5,\,d_2=2,\,q=2$.} 
 \end{center}
\end{figure}
\begin{figure}
 \begin{center}
  \includegraphics[width=6.3truecm,clip]{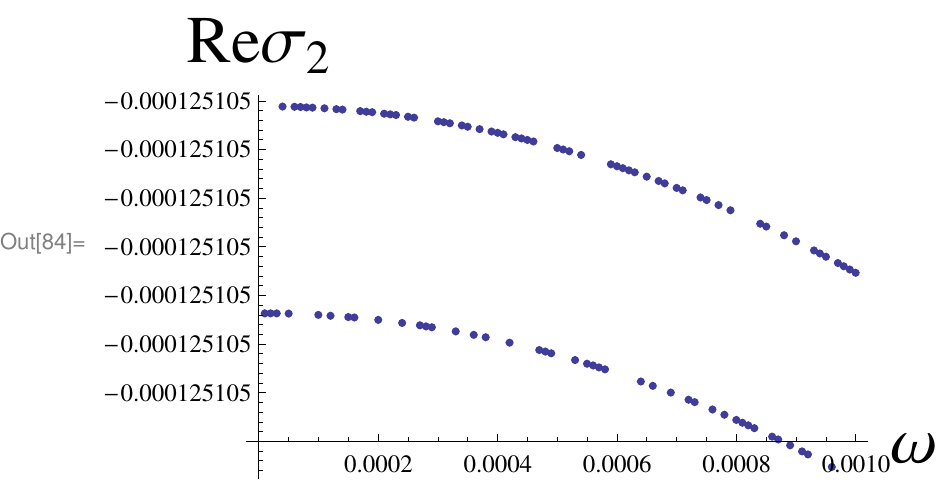}
  \caption{(color online) $\mbox{Re}[\overline{\sigma(\omega)^{(2)}}]$ in the case of 
$d_1=5,\,d_2=2,\,q=5$.} 
 \end{center}
\end{figure}

Note that even though Re[$\sigma(\omega)^{(i)}$] $< 0$ ($i=1,2$), since we have 
small but nonzero Re[$\sigma(\omega)^{(0)}$] $>0$, as long as we consider the 
perturbative analysis, 
the real part of the conductivities are always positive.


\end{document}